\documentclass[fleqn,usenatbib]{mnras}

\usepackage{newtxtext,newtxmath}

\usepackage[T1]{fontenc}

\DeclareRobustCommand{\VAN}[3]{#2}
\let\VANthebibliography\thebibliography
\def\thebibliography{\DeclareRobustCommand{\VAN}[3]{##3}\VANthebibliography}

\usepackage{graphicx}	
\usepackage{amsmath}	

\title[Inferring stellar mass growth histories]{Hierarchical assembly impedes the inference of stellar mass growth histories for individual galaxies}

\author[R. K. Cochrane]{
R. K. Cochrane$^{1,2}$\thanks{E-mail: rachel.cochrane@manchester.ac.uk}
\\
$^{1}$Jodrell Bank Centre for Astrophysics, University of Manchester, Oxford Road, Manchester M13 9PL, UK\\
$^{2}$Institute for Astronomy, University of Edinburgh, Royal Observatory, Blackford Hill, Edinburgh, EH9 3HJ, UK}
\date{Accepted 2025 September 2. Received 2025 August 29; in original form 2025 August 3}
\pubyear{\the\year{}}
\begin{document}
\label{firstpage}
\pagerange{\pageref{firstpage}--\pageref{lastpage}}
\maketitle

\begin{abstract}
Some massive, quiescent galaxies at $z>3$ appear to contain considerable numbers of old stars (forming at $z>7$). Works inferring the star formation histories of at least one such galaxy observed with JWST have suggested that the assembly of so much stellar mass so early may challenge the well-established Cold Dark Matter (CDM) cosmological framework, or else indicate extraordinarily high past star formation efficiencies. However, these studies implicitly assume that all the stars in place at the epoch of observation assembled in-situ, i.e. in a single galaxy. In hierarchical models like CDM, massive galaxies assemble following successive mergers of smaller galaxies. Thus, inferences of the growth of stellar mass using the ages of stars within a descendant massive galaxy will be biased. In this paper, I use the TNG100 simulation to quantify this bias across a range of descendant galaxy masses and redshifts, for inferences made for different past epochs. I demonstrate that the assumption of in-situ stellar mass assembly can lead to significant biases in inferred stellar mass histories, with historic (i.e. looking back from the epoch of observation) stellar masses overestimated by over an order of magnitude in some cases. The bias increases with increasing halo mass, for inferences made further from the epoch of observation, and with decreasing descendant galaxy redshift. I derive corrections that can be applied to inferred stellar mass histories for more robust comparisons with dark matter halo mass functions at high redshift. \vspace{0.2cm}
\end{abstract}




\section{Introduction}
Contemporary studies of galaxy formation are built upon the foundations of the cold dark matter (CDM) cosmological framework \citep{Peebles1982,Blumenthal1984,Davis1985}. In hierarchical models like CDM, the assembly of structure proceeds via the aggregation of smaller structures \citep{Peebles1970}, with peaks in the density field of dark matter after recombination evolving under gravity into dark matter halos with a distribution of masses \citep{Press1974,Peacock1990,Bond1991,Bower1991}. Gas cools radiatively to form galaxies in the cores of massive halos \citep{White1978}. Following mergers of dark matter halos, the rates of which can be calculated analytically for halos of different masses \citep[e.g.][]{Lacey1993,Lacey1994}, their constituent galaxies eventually merge under dynamical friction, building more massive galaxies. \\
\indent The early formation times inferred from observations of some massive galaxies \citep[e.g.][]{Brinchmann2000,Glazebrook2004,vanderWel2005,Bundy2006} have, in the past, been invoked as a challenge to CDM \cite[see][for a review]{Renzini2006}. Historically, semi-analytic models struggled to match both local luminosity functions and the observed number densities of massive galaxies at $z>1$ \citep[e.g.][]{Baugh1998,Baugh2005,Kauffmann1999,Kauffmann1999a}. `Downsizing', i.e. the most massive galaxies forming earliest and the luminosities of the brightest objects decreasing towards low redshift \citep{Cowie1996}, at first appeared in tension with the hierarchical nature of CDM. Observational data were later shown to be explicable within the CDM paradigm using more sophisticated models for baryonic physics, in particular by invoking AGN feedback and modelling its impact on star formation efficiency at different epochs \citep{Croton2005,DeLucia2006,Bower2006}. As outlined by \cite{Bower2006}, for a given halo mass, the ratio of gas cooling time to free-fall time is lower at higher redshift. Black holes are also less massive at higher redshift. For these two reasons, AGN feedback is less effective for high mass halos at high-redshift, enabling more rapid stellar mass growth; at lower redshifts (particularly at $z\lesssim1$), AGN feedback suppresses cooling flows and prevents the growth of the most massive galaxies. \\
\indent The sensitivity and spectral coverage of JWST has enabled deep, rest-frame optical spectra (importantly, constraining the Balmer break) to be obtained for tens of massive quiescent galaxy candidates at $z\gtrsim3$ \citep{Carnall2023,Carnall2023a,Carnall2024,DeGraaff2024,Glazebrook2024,Nanayakkara2024}. This enables physical characterisation of fainter and older quiescent galaxies than was possible with ground-based facilities.
Some of these studies have again suggested that observations of high-redshift massive galaxies challenge hierarchical models.
\cite{Glazebrook2024} presented observations of a
previously-identified quiescent candidate \citep[ZF-UDS-7329;][]{Schreiber2018} with JWST NIRSpec's PRISM disperser ($R=50-350$ from $\lambda=0.6-5.0\,\mu\rm{m}$, with increasing resolution at longer wavelengths). Fitting the spectrum and photometry with the FAST++ Spectral Energy Distribution (SED) fitting code \citep{2018ascl.soft03008K}, they derived a stellar mass of $\log_{10}(M_{\star}/\rm{M_{\odot}})=11.26^{+0.03}_{-0.16}$ for this $z=3.205$ source. Crucially, the fit suggested a very early formation time, with half of the stellar mass formed by $1.52\pm0.16\,\rm{Gyr}$ prior to the epoch of observation, i.e. by $z\sim10$. \cite{Glazebrook2024} concluded that such early formation of such a massive galaxy might threaten $\Lambda\rm{CDM}$. Re-observing the same source at higher spectral resolution, \cite{Carnall2024} fitted the new spectrum and drew broadly consistent conclusions: a stellar mass of $\log_{10}(M_{\star}/\rm{M_{\odot}})=11.14\pm0.03$, with $50\%$ of the $z=3.205$ stellar mass formed by $z=10.4^{+4.0}_{-2.2}$. They derived $M_{\star}$ vs $z$ by integrating the fitted star formation history, then compared this stellar assembly history to the stellar mass expected for the most massive halo within the appropriate observed volume, given different assumptions for star formation efficiencies \cite[see][for details]{Lovell2023}. From this comparison, \cite{Carnall2024} concluded that the stellar mass inferred for this galaxy at $z\sim7-9$ requires extremely efficient ($100\%$) conversion of baryons into stars. \cite{DeGraaff2024} derived a similar formation time for a source at even higher redshift ($z=4.9$), but cautioned that the inferred formation time is degenerate with metallicity. \cite{Turner2025} also highlighted the dependence of inferred formation time on the prior assumption on the form of the star formation history (SFH). Refitting the spectrum of ZF-UDS-7329 obtained by \cite{Glazebrook2024}, they showed that a flat SFH prior yielded a formation time consistent with that derived by \cite{Glazebrook2024}, but that assuming a rising SFH prior decreased this to $z\sim8$. Hence, some of the apparent tension with $\Lambda\rm{CDM}$ could be eased by different fitting assumptions.\\
\indent One source of uncertainty that has not been examined in these studies is how hierarchical assembly naturally poses a challenge for stellar mass assembly history inference. A perfect SED fitting code would yield ages of all stars within a galaxy at the time of observation. However, since massive galaxies assemble via a combination of in-situ and ex-situ star formation, it is not clear that such ages can be used to reconstruct the stellar mass history of an individual massive galaxy progenitor, as has been attempted by \cite{Glazebrook2024} and \cite{Carnall2024}. The technique used by these authors is strictly only valid in the case that the observed galaxy has only a single massive progenitor. This is of particular concern for inferences of stellar mass assembly histories at very early epochs, since merger rates were higher at high redshifts for both halo-halo and galaxy-galaxy mergers \citep{Rodriguez-Gomez2015,Duan2025,Puskas2025}. \cite{Turner2025} demonstrated the impact of one or more major mergers of ZF-UDS-7329 at $z=6$ on the inferred star formation efficiency required at earlier times, also noting the need to consider cosmic variance \citep[see also][]{Jespersen2025}. However, the impact of multiple progenitors on the derived stellar mass assembly histories of massive galaxies has not been explored in detail. \\
\indent The aim of this paper is to quantify how important these effects are and whether they need to be considered when inferring the stellar mass histories of observed galaxies. 
I use public data from The Next Generation Illustris project\footnote{\url{https://www.tng-project.org/data/}}, hereafter IllustrisTNG, to track the stellar assembly of galaxies across a range of stellar masses. I compare the stellar mass formed in the most massive progenitor of a given galaxy to that inferred from the ages of its stars. The structure of this paper is as follows. In Section \ref{sec:methods}, I describe the simulations and the various methods used to construct stellar mass assembly histories. In Section \ref{sec:results}, I quantify the bias in inferred stellar mass assembly history that results from assuming a single progenitor. I provide average correction factors that can be applied to observationally-inferred stellar mass growth histories. In Section \ref{sec:discussion}, I explore the implications of this work for observational studies of high-redshift galaxies. I draw conclusions in Section \ref{sec:conclusions}.\\
\indent For consistency, I adopt the cosmology used by IllustrisTNG: $\Omega_{\Lambda,0}=0.6911$, $\Omega_{\rm{m},0}=0.3089$, $\Omega_{\rm{b},0}=0.0486$, $\sigma_{8}=0.8159$, $n_{s}=0.9667$, and $h=0.6774$ \citep{PlanckCollaboration2015}.

\section{Methods}\label{sec:methods}
\subsection{TNG100 simulation}
The IllustrisTNG simulation suite \citep{Marinacci2018,Springel2018,Nelson2018a,Pillepich2018,Naiman2018} represents the state-of-the-art in large-scale cosmological hydrodynamical simulations. IllustrisTNG is the latest addition to the Illustris Project \citep{Vogelsberger2014a,Vogelsberger2014,Genel2014,Sijacki2015}. The IllustrisTNG suite consists of three main simulation volumes:
TNG50, TNG100, and TNG300. I choose an appropriate volume based on balancing the need for samples of massive, high-redshift galaxies (which demands a large box) with the need for sufficient particle resolution to model the lower mass progenitors of these galaxies at earlier epochs. TNG100 (box side length $100\,\rm{comoving\,\,Mpc}$) satisfies both requirements. In TNG100, dark matter particles have a mass of $7.46\times10^{6}\,\rm{M_{\odot}}$ and baryons (i.e. star particles and gas cells) have a mass of $1.39\times10^{6}\,\rm{M_{\odot}}$. The mass resolution of TNG300 is almost $10$ times lower. Gravitational softening lengths in TNG100 are $\sim0.7\,\rm{kpc}$ for dark matter and star particles. The softening of the gas cells is adaptive, with a minimum value of $\sim0.18\,\rm{kpc}$. I provide details of the resolution tests I have performed to confirm convergence in Appendix \ref{sec:appendix_res_tests}.
IllustrisTNG simulations are run with {\sc{arepo}} \citep{Springel2010,Weinberger2020}, a massively parallel code that solves the coupled equations of gravity and magnetohydrodynamics across a large dynamic range. The implemented models for galaxy formation physics, including radiative gas cooling, star formation physics, models for supernova-driven galactic winds, metal enrichment, and the growth of and feedback from supermassive black holes, are described in detail in \cite{Pillepich2018}. 

\begin{figure*}
    \centering
    \includegraphics[width=1.0\linewidth]{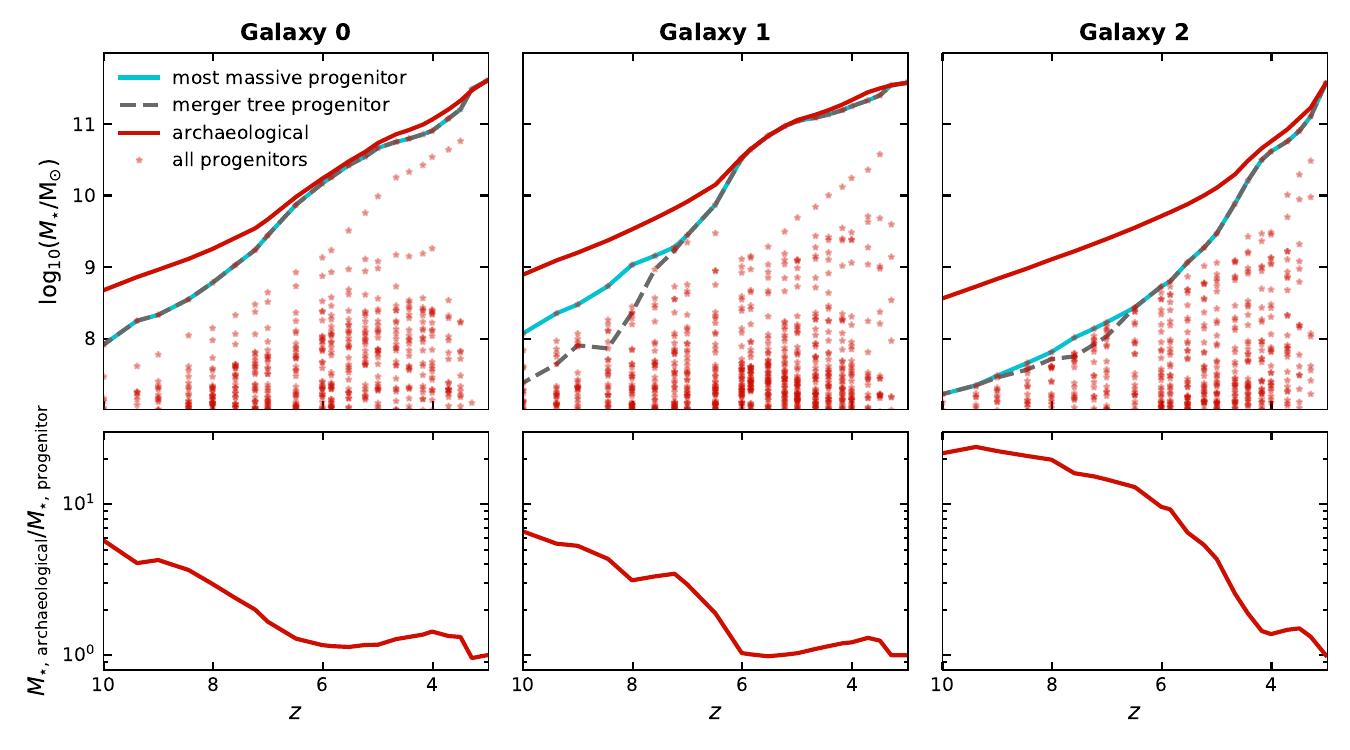}
    \vspace{-0.8cm}
    \caption{Upper panels: stellar mass assembly histories for the three most massive galaxies in the TNG100 simulation at $z=3$, which all exceed $M_{\star}=10^{11}\,\rm{M_{\odot}}$ by this epoch. Solid red lines show the stellar mass history constructed from the ages of stars within a massive galaxy at $z=3$ (the `archaeological' approach). Blue lines show the stellar mass history constructed by tracking the stars within the galaxy's most massive progenitor at earlier cosmic times; grey dashed lines show the main progenitor identified using the {\sc sublink} algorithm. The masses of all progenitor galaxies with $M_{\star}>10^{7}\,\rm{M_{\odot}}$ are marked with red stars. Lower panels: the ratio of the archaeologically-inferred and true stellar mass within the most massive progenitor (blue dotted lines on the upper panel), as a function of redshift. The archaeological approach leads to an overestimate of the stellar mass within the galaxy at early times (by over an order of magnitude, in some cases), due to the incorrect assumption that all the stellar mass present at $z=3$ was formed in the same dark matter subhalo.}
    \label{fig:halo_hist}
\end{figure*}

\subsection{Star formation histories from halo merger trees}
Here, I summarise the steps taken to construct star formation histories for massive galaxies in TNG100. I make use of publicly-available subhalo merger trees, which link dark matter halo/galaxy progenitors and their descendants across simulation snapshots. As described by \cite{Rodriguez-Gomez2015}, dark matter halos are identified using a friends-of-friends (FoF) approach \citep{Davis1985}, and baryonic resolution elements are assigned to FoF groups based on their nearest dark matter particle. An algorithm based on {\sc subfind} \citep{Springel2001,Dolag2009} is used to identify gravitationally-bound substructures within the FoF groups. These are then linked through cosmic time using the {\sc sublink} algorithm \cite[see][for a comprehensive study of different tree-building algorithms]{Srisawat2013,Avila2014,Lee2014}, following the procedure described in \cite{Rodriguez-Gomez2015}. Each subhalo is assigned a unique descendant based on subhalos in the next snapshot with particles in common. Each subhalo can have many progenitors but a maximum of one descendant. The first progenitor of each subhalo is defined as a the one with the `most massive history' behind it, i.e. the branch that accounts for the most of the mass of the descendant for the longest time period, following \cite{DeLucia2007}. \\
\indent In Figure \ref{fig:halo_hist} (upper panels), I show examples of the progenitors of the three most massive galaxies in the TNG100 catalogue at $z=3$. As expected, each `descendant' $z=3$ galaxy assembles from many progenitors (see small red stars). The stellar mass assembly of the {\sc sublink}-identified first progenitor is shown in grey (dashed lines). I note that, at some redshifts, this first progenitor is not the most massive of all progenitors (see snapshots at $z\gtrsim7$ for Galaxy 1 and $6.5\lesssim z \lesssim8$ for Galaxy 2). This appears to occur when there is a merger between two galaxies of similar stellar masses. Using the first progenitor history would fail to capture the most massive progenitor at some epochs, for some galaxies. The aim of this paper is to quantify the discrepancy between the most massive progenitor at early times and the stellar mass of the progenitor inferred from SED fitting. Hence, for every descendant at its `final' redshift, I identify the single most massive progenitor at each epoch. This sequence is plotted for the three examples in blue. For the rest of this paper, I adopt this alternative `most massive progenitor' stellar mass assembly history.

\subsection{Reconstructing archaeological star formation histories}
In the ideal case, SED fitting yields ages for all stars in the observed galaxy, from which a stellar mass assembly history can be constructed. For each descendant TNG100 galaxy, I generate this `archaeological' stellar assembly history from the ages of its surviving stars. Examples are shown in Figure \ref{fig:halo_hist} (upper panels, solid red lines). Near the time of observation, these archaeological assembly histories are similar to the `most massive progenitor' tracks; at these epochs, most of the mass within the descendant is contained within just one massive progenitor. However, at higher redshifts, the two tracks diverge, and the archaeological stellar assembly history exceeds that of the single most massive progenitor. This arises because more than one progenitor contributes significant numbers of old stars to the descendant. At these epochs, even perfect observational inference of the formation histories of stars within a galaxy will result in a significant overestimation of the stellar mass within the single most massive progenitor. I quantify $M_{\star,\,\rm{archaeological}}/M_{\star,\,\rm{progenitor}}$, as a function of redshift, for each of the three massive galaxies at $z=3$. {\it{This can be considered as the bias in the stellar mass history inferred from SED fitting (if the stellar ages of all stars in the descendant galaxy were recovered perfectly), compared to the true stellar mass of the most massive progenitor at each epoch.}}\footnote{We do not present tests of the SFH inference in this paper. Future work could explore the fidelity of the SFH recovery of massive quenched galaxies, for different types of data, e.g. photometry versus spectra of different quality.} I will use this ratio to characterise bias in the inferred stellar mass growth history of individual galaxies throughout this paper. I show examples in the lower panels of Figure \ref{fig:halo_hist}. Importantly, this bias can exceed a factor of $10$ at early times (see Galaxy 2 at $z\gtrsim6$). In the following Section, I quantify this bias as a function of descendant galaxy properties and redshift. 

\begin{figure*}
    \centering    
    \includegraphics[width=1.0\linewidth]{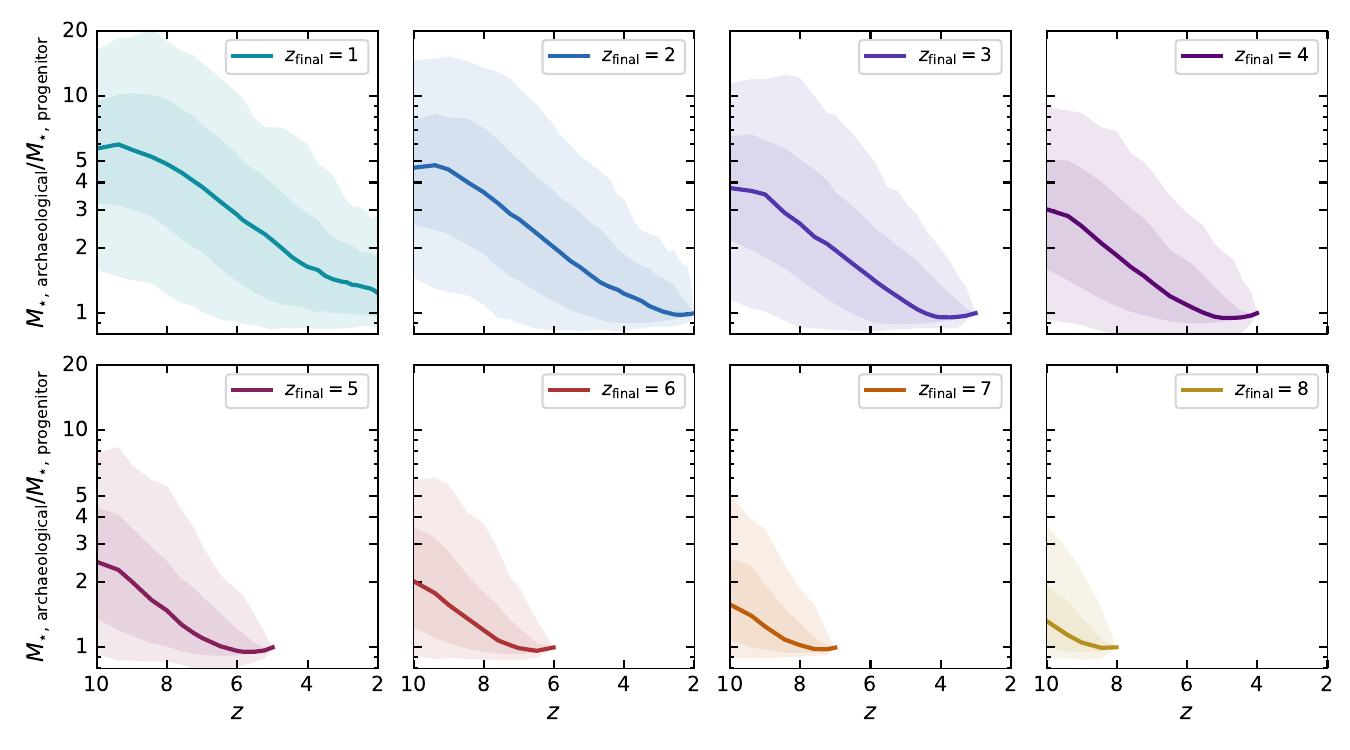}
    \vspace{-0.7cm}
    \caption{Ratio of the archaeologically-inferred to true most-massive progenitor stellar mass versus redshift, constructed using the $500$ most massive galaxies in TNG100 at final `observed' redshifts between $z=8$ and $z=1$ (see legends). Solid lines show the $50^{\rm{th}}$ percentile of the ratio at each redshift. Darker (lighter) filled regions show the $16^{\rm{th}}-84^{\rm{th}}$ ($2^{\rm{nd}}-98^{\rm{th}}$) percentile intervals. The bias in inferred stellar mass is greater for sources observed at lower redshifts, for which inferences are made further from the time of observation and for which there is more cosmic time to undergo mergers. The median stellar masses of the subsamples at $z=1,2,3,4,5,6,7,\,\rm{and}\,\,8$ are $\log_{10}M_{\star}/\rm{M_{\odot}}=11.1,10.9,10.6,10.2,9.7,9.1,8.7\,\rm{and}\,\,8.3$, respectively.}
    \label{fig:500_most_massive}
\end{figure*}

\begin{figure}
    \centering
    \includegraphics[width=1.0\linewidth]{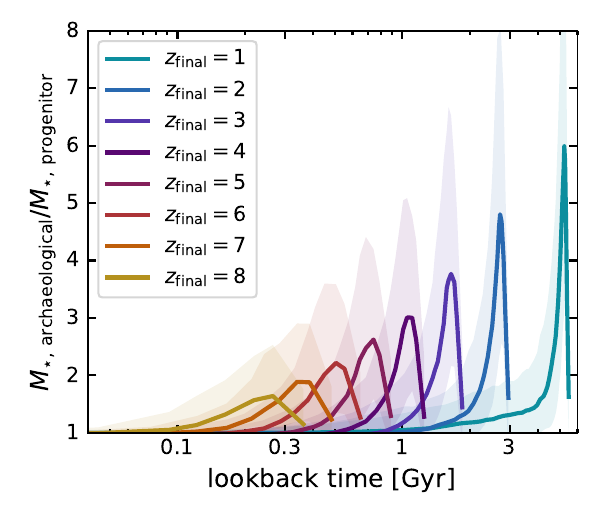}
    \vspace{-0.7cm}
    \caption{Ratio of the archaeologically-inferred to true most-massive progenitor stellar mass ratio versus lookback time, for descendant massive galaxies at $z=8-1$. This ratio increases more rapidly with lookback time for galaxies that are observed at higher redshifts; here, the merger rate is higher close to the epoch of observation. However, the ratio reaches greater values, on average, for galaxies observed at lower redshifts, which have more cosmic time to experience mergers. As for Figure \ref{fig:500_most_massive}, I construct this figure using the 500 most massive descendants at each final redshift. For clarity of the figure, I show only the median relation (solid coloured lines) and $16^{\rm{th}}-84^{\rm{th}}$ percentile interval (shaded regions).}\label{fig:ratio_vs_lookbacktime}
\end{figure}

\section{Results}\label{sec:results}
\subsection{The overestimation of historic stellar mass, as a function of redshift of observation and lookback time}
In this section, I explore trends in $M_{\star,\,\rm{archaeological}}/M_{\star,\,\rm{progenitor}}$ across redshift. I calculate $M_{\star,\,\rm{archaeological}}/M_{\star,\,\rm{progenitor}}$ versus $z$ for the $500$ most massive descendant galaxies at each of $z=1$, $z=2$, $z=3$, $z=4$, $z=5$, $z=6$, $z=7$, and $z=8$. In Figure \ref{fig:500_most_massive}, I present distributions of these ratios versus progenitor redshift. On each panel, the solid line shows the $50^{\rm{th}}$ percentile of $M_{\star,\,\rm{archaeological}}/M_{\star,\,\rm{progenitor}}$ at each epoch. I also mark the $16^{\rm{th}}-84^{\rm{th}}$ and $2^{\rm{nd}}-98^{\rm{th}}$ percentiles. The large span of these intervals at any given lookback time reflects significant galaxy-to-galaxy differences. Nevertheless, we can note some general trends. As is expected from hierarchical assembly, $M_{\star,\,\rm{archaeological}}/M_{\star,\,\rm{progenitor}}$ increases with lookback time, i.e. constructing stellar mass histories becomes harder further from the epoch of observation. This is because by the epoch of observation, lower-redshift galaxies have had more time to undergo galaxy mergers, and hence will be comprised of more individual progenitors. For example, for a massive galaxy observed at $z=1$ ($z=2$), the stellar mass of an individual progenitor is, on average, overestimated by a factor of $\gtrsim3$ ($\gtrsim2$) above $z=6$. It is interesting that these biases can be so large at early times, even when the overall ex-situ mass within the descendant is low;
this is because the bulk of the stellar mass of these simulated galaxies forms in-situ closer to the epoch of observation (see Figure \ref{fig:halo_hist}: the stellar mass of all three massive galaxies increases by a factor of $>10$ between $z=6$ and $z=3$). I display the same data in a different way in Figure \ref{fig:ratio_vs_lookbacktime}. Here, I show $M_{\star,\,\rm{archaeological}}/M_{\star,\,\rm{progenitor}}$ as a function of lookback time (measured from the time of observation) for descendants at all eight final redshifts. As was clear from Figure \ref{fig:500_most_massive}, $M_{\star,\,\rm{archaeological}}/M_{\star,\,\rm{progenitor}}$ increases with lookback time, with the largest values seen for galaxies observed at lower redshifts. In addition, the change in $M_{\star,\,\rm{archaeological}}/M_{\star,\,\rm{progenitor}}$ with lookback time can be seen more clearly in this figure. $M_{\star,\,\rm{archaeological}}/M_{\star,\,\rm{progenitor}}$ increases most rapidly with lookback time for descendant galaxies at high redshifts. Here, the merger rate is higher closer to the epoch of observation.

\subsection{The overestimation of historic stellar mass, as a function of galaxy/halo properties}\label{sec:gal_property_dependence}
\begin{figure}
\includegraphics[width=0.9\linewidth]{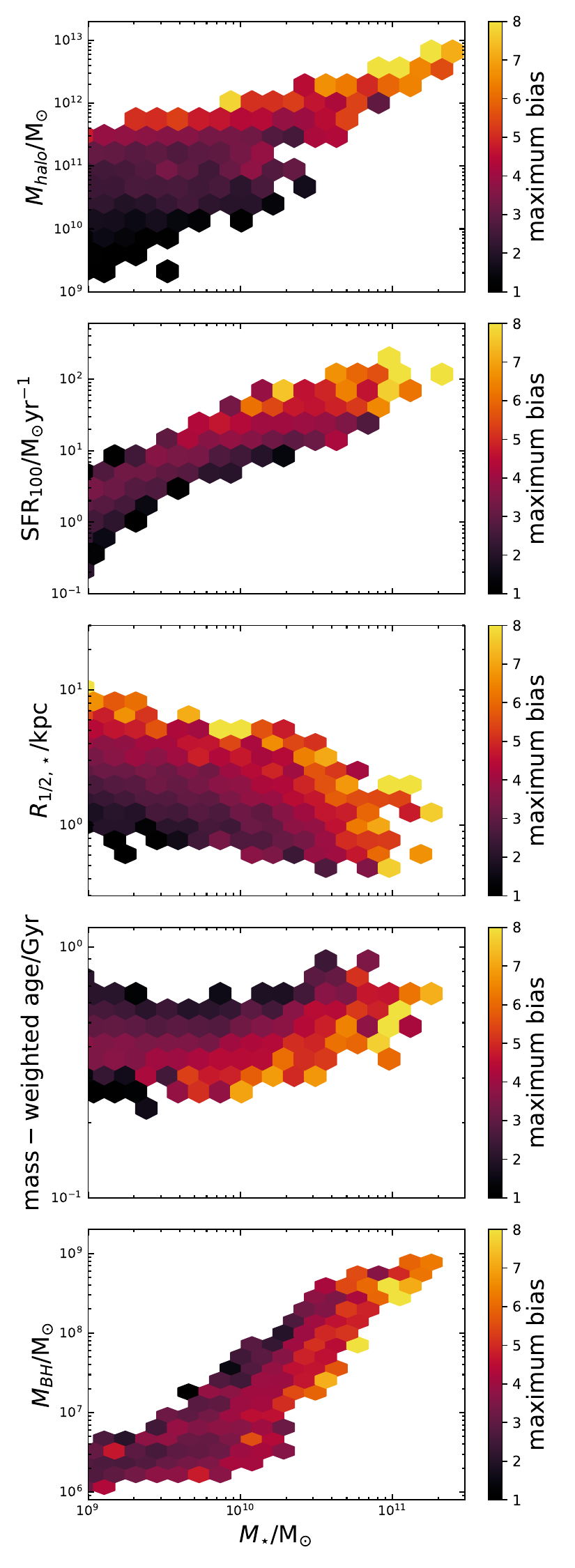}
\vspace{-0.5cm}
\caption{Key galaxy scaling relations (from top to bottom: halo mass, SFR, half-mass radius, mass-weighted age and black hole mass versus stellar mass), for TNG100 galaxies at $z=3$. All panels are colour-coded by the median maximum $M_{\star,\,\rm{archaeological}}/M_{\star,\,\rm{progenitor}}$ (the `maximum bias') reached by the galaxies within each bin. The highest maximum bias values occur for galaxies with high halo masses, which also tend to have the highest stellar masses, SFRs and black hole masses. In Figure \ref{fig:separate_halomass}, I show that halo mass is the key parameter correlating with maximal bias.}
\label{fig:hexbins}
\end{figure}

Here, I investigate correlations between the bias ($M_{\star,\,\rm{archaeological}}/M_{\star,\,\rm{progenitor}}$) and galaxy properties. For each descendant galaxy, I calculate the maximum value of $M_{\star,\,\rm{archaeological}}/M_{\star,\,\rm{progenitor}}$ (the `maximum bias') reached at any redshift. I then look for trends in this maximum bias with key galaxy/halo properties of the descendant: host halo mass, host stellar mass, SFR, half-mass radius, mass-weighted stellar age, and black hole mass. I present these data for the $z=3$ descendant sample in Figure \ref{fig:hexbins}. The median maximum bias reached for galaxies within each hexagonal bin (see colour bar) correlates with various galaxy properties. The highest maximum bias is seen at high halo mass and high stellar mass, with some weak trends also seen with half-mass radius, SFR and mass-weighted age. However, since these galaxy parameters are all correlated, it is difficult to identify which parameters are most fundamentally related to the bias. In Figure \ref{fig:separate_halomass}, I plot maximum bias versus each of the galaxy properties, in bins of halo mass. There is very little remaining dependence of the maximum bias with any galaxy property, once the halo mass dependence is removed. This indicates that halo mass is the fundamental parameter that sets the bias. I show these figures for a single redshift ($z=3$), to match the most interesting redshift range of observed samples of massive, quiescent galaxies. However, a consistent picture is also seen at other redshifts.

\subsection{Simulation-derived corrections for stellar mass histories}
In Section \ref{sec:gal_property_dependence}, I identified host halo mass as the parameter most strongly linked to the maximum value of $M_{\star,\,\rm{archaeological}}/M_{\star,\,\rm{progenitor}}$ reached at any point in a galaxy's history. Although halo mass is difficult to measure for individual galaxies, it is strongly correlated with stellar mass, which is a much more accessible property to measure. Here, I derive distributions of $M_{\star,\,\rm{archaeological}}/M_{\star,\,\rm{progenitor}}$ as a function of descendant galaxy stellar mass and redshift. \\
\indent In Figure \ref{fig:mass_dependence}, I show $M_{\star,\,\rm{archaeological}}/M_{\star,\,\rm{progenitor}}$ versus stellar mass at different epochs in the history of descendant massive galaxies at $z=5$ (left), $z=3$ (centre) and $z=1$ (right). I note two significant trends. Firstly, at fixed descendant stellar mass, stellar mass histories inferred using the archaeological approach further from the epoch of observation are generally most biased. This is in line with the previous figures and interpretation. For a descendant galaxy of stellar mass $10^{11}\,\rm{M_{\odot}}$ at $z=1$, $M_{\star,\,\rm{archaeological}}/M_{\star,\,\rm{progenitor}}$ increases from $\sim2$ at $z=5$ to $\sim4$ at $z=7$. Secondly, stellar mass histories inferred for more massive descendant galaxies are generally more biased than for less massive galaxies. This trend is seen most clearly for the $z=1$ descendant galaxies (right-hand panel), which reach the highest stellar masses. This can be understood in the context of the halo-to-stellar mass relation. As shown in Figure \ref{fig:halo_to_stellar_mass}, the relation between halo mass and stellar mass steepens above $M_{\star} \sim4\times10^{10}\,\rm{M_{\odot}}$. This, in combination with the increase in bias with halo mass, drives a steepening relation between bias and stellar mass at high stellar masses.\\
\indent In supplementary data released alongside the published version of this paper, I will provide the data displayed in Figure \ref{fig:mass_dependence} alongside example code. This will enable simulations-motivated approximate corrections to be applied to inferred stellar mass histories. Given the dependence of the identified bias on galaxy redshift and host halo mass, ideally studies should derive more robust corrections matched to their observed sample.

\section{Discussion}\label{sec:discussion}
In this section, I summarise this work and discuss implications for observational studies. As discussed in the Introduction, JWST is providing unparalleled spectroscopic data probing rest-optical wavelengths for massive galaxies at $z\gtrsim3$. Some of the most interesting targets are massive, quiescent candidates that appear to have stopped forming stars well before the peak of cosmic star formation, at $z\sim3-5$ \citep{Carnall2023,Carnall2023a,Carnall2024,Glazebrook2024,Nanayakkara2024a,Nanayakkara2024,Setton2024a,Antwi-Danso2025}. These new data, probing some of the most massive galaxies in the early Universe, which likely form within particularly overdense environments, have motivated comparisons with galaxy formation models. Spectral fitting suggests that the stars within these galaxies formed extremely early, with half of the stellar mass of some galaxies formed prior to $z\sim10$ \citep{Glazebrook2024,Carnall2024,DeGraaff2024}. To set these galaxies in a cosmological context, these works have compared the stellar mass histories constructed from the inferred star formation histories to maximal halo masses expected given the survey volume. They suggested that high stellar masses at $z\gtrsim7$ challenge the well-established $\Lambda\rm{CDM}$ model and/or require extremely high star formation efficiencies. However, it is not clear that such comparisons with dark matter halo growth models are robust. The constructed stellar mass histories assume that all the stellar mass in the observed galaxy was formed in-situ, rather than in multiple progenitors that have undergone mergers to form the massive galaxy. I explore the impacts of that assumption in this paper. 

\begin{figure}
\includegraphics[width=0.9\linewidth]{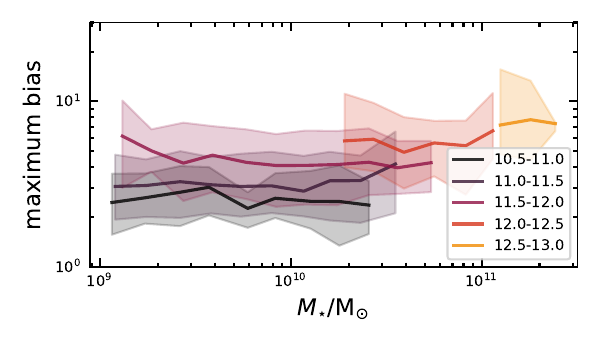}\vspace{-0.4cm}
\includegraphics[width=0.9\linewidth]{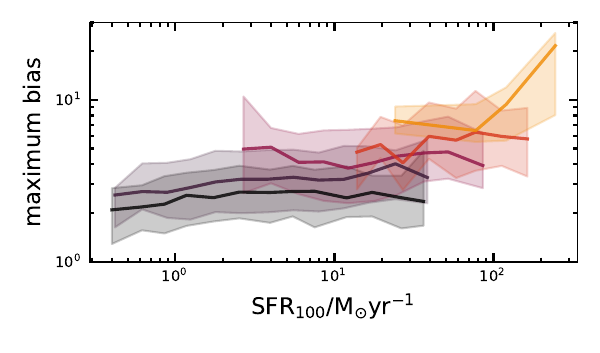}\vspace{-0.4cm}
\includegraphics[width=0.9\linewidth]{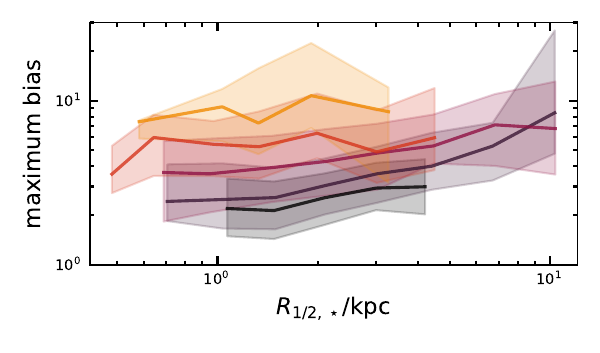}\vspace{-0.4cm}
\includegraphics[width=0.9\linewidth]{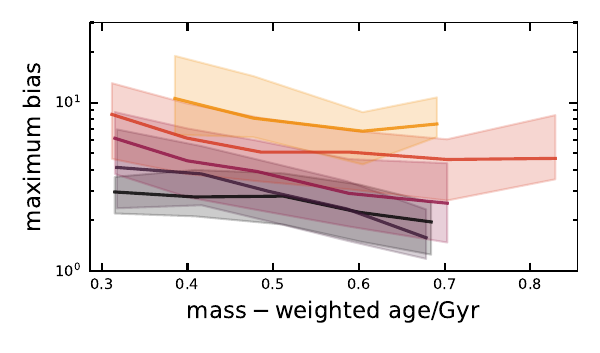}\vspace{-0.4cm}
\includegraphics[width=0.9\linewidth]{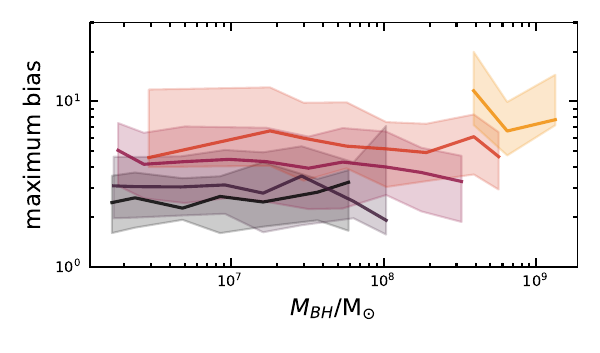}\vspace{-0.5cm}
\caption{The maximum value of $M_{\star,\,\rm{archaeological}}/M_{\star,\,\rm{progenitor}}$ (the maximum `bias') reached versus each galaxy property, for galaxies at $z=3$ binned by halo mass (see legend for $\log_{10}(M_{\rm{halo}}/\rm{M_{\odot}})$ range). At a given halo mass, there is little dependence on any galaxy property.}
\label{fig:separate_halomass}
\end{figure}

\begin{figure*}
    \centering
    \includegraphics[width=0.33\linewidth]{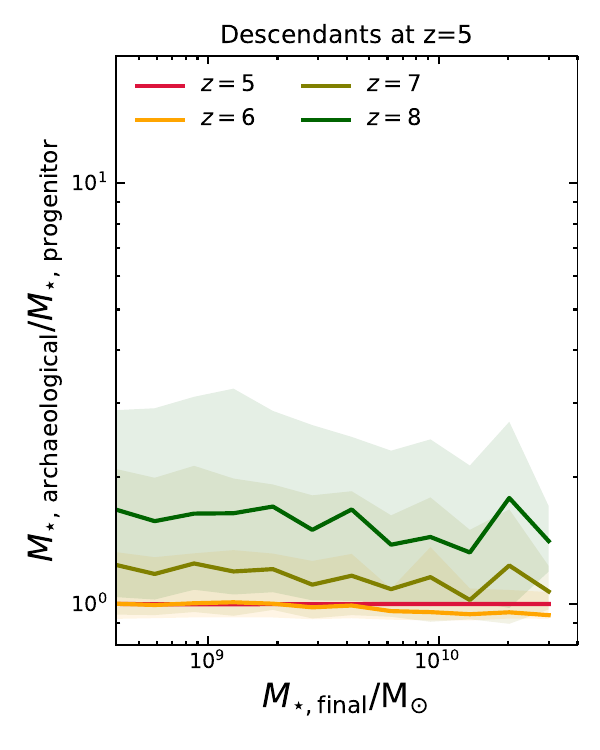}
    \includegraphics[width=0.33\linewidth]{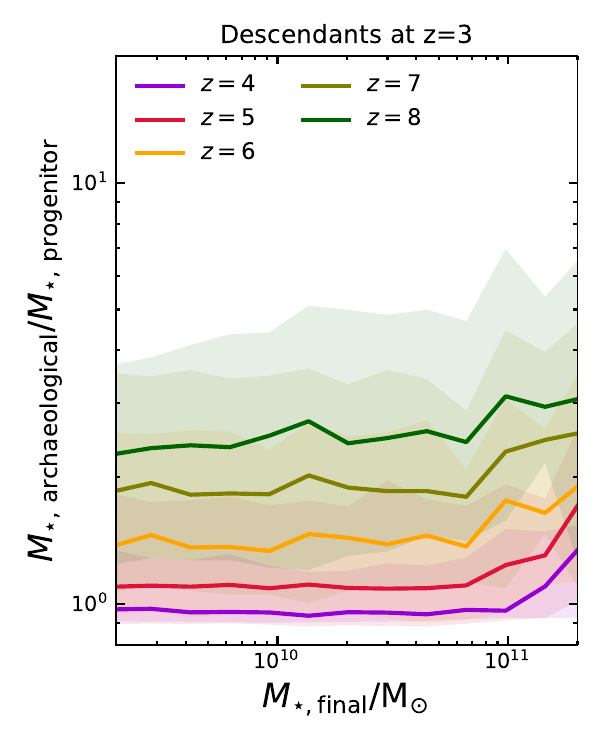}
    \includegraphics[width=0.33\linewidth]{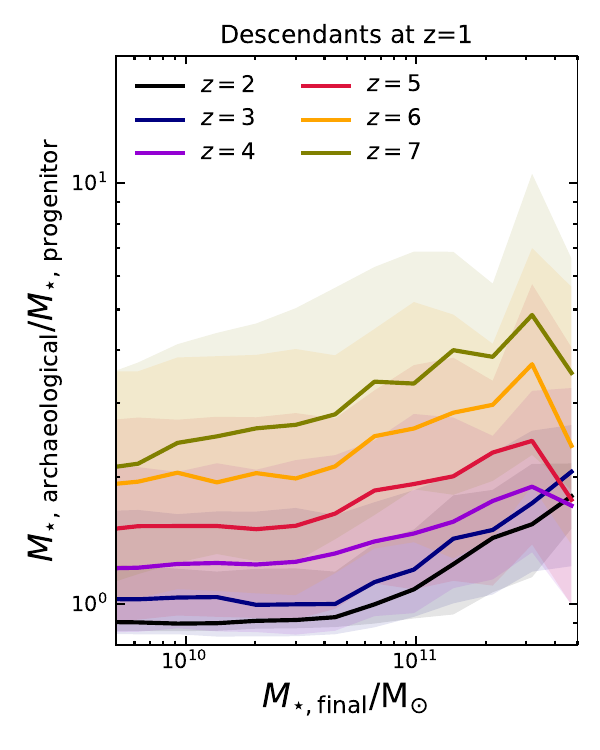}
    \vspace{-0.5cm}
    \caption{Ratio of the archaeologically-inferred to true progenitor stellar mass (i.e. the bias in inferred historic stellar mass) at different redshifts, as a function of final stellar mass, constructed using $\gtrsim3000$ massive galaxies at each of $z=5$ (left), $z=3$ (centre), and $z=1$ (right). Solid lines show the $50^{\rm{th}}$ percentile of $M_{\star,\,\rm{archaeological}}/M_{\star,\,\rm{progenitor}}$ within each descendant stellar mass bin. Shaded regions show the $16^{\rm{th}}-84^{\rm{th}}$ percentile interval. Stellar masses inferred using the archaeological approach at higher redshifts (i.e. longer from the time of observation) tend to be more biased than at lower redshifts. There is also a shallow dependence on final stellar mass, with the most massive galaxies being most affected, on average. This stellar mass dependence is clearer for lower redshift descendants, which naturally reach higher final stellar masses.}
    \label{fig:mass_dependence}
\end{figure*}

\begin{figure}
    \centering    \includegraphics[width=\linewidth]{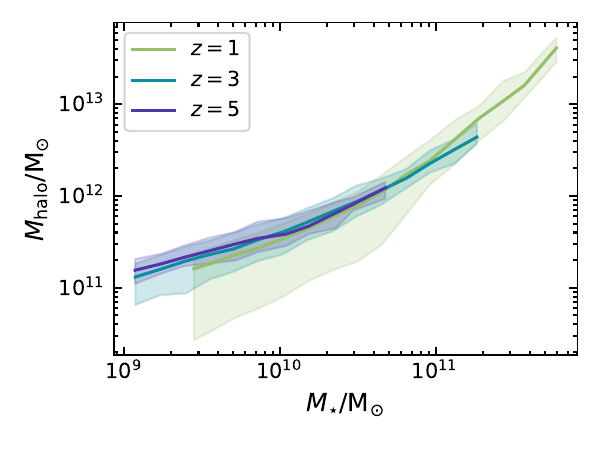}    \vspace{-0.7cm}
    \caption{The median (solid lines) and $1\sigma$ range (shaded regions) of halo mass spanned at a given stellar mass, for galaxies at $z=1-3$. The halo-to-stellar mass relation steepens above $\sim4\times10^{10}\,\rm{M_{\odot}}$. This drives the steepening trend in bias versus stellar mass seen at the highest stellar masses (see Figure \ref{fig:mass_dependence}), which are only reached by galaxies at the lowest redshifts studied in this paper.}
\label{fig:halo_to_stellar_mass}
\end{figure}

\subsection{Key results in this paper}
In this paper, I quantified the importance of ex-situ assembly on the stellar mass history, and hence characterised the bias in stellar mass histories that can be inferred from spectroscopic observations. Simulations, which track the assembly of halos and galaxies through successive mergers, provide an ideal test-bed for this study. I selected massive `descendant' galaxies at $z=1-8$ from the large-box TNG100 cosmological simulation and extracted their detailed assembly histories using subhalo merger trees. I demonstrated that stellar mass histories reconstructed from star formation histories at the epoch of observation can differ from the stellar mass history of the most massive individual progenitor by over an order of magnitude at some epochs (e.g. see Figure \ref{fig:halo_hist}, right-hand panel, where the archaeological stellar mass exceeds that of the most massive progenitor by a factor of $>20$ at $z\gtrsim8$). This is true even if the inferred star formation history is {\it perfect}, and is a predictable consequence of hierarchical galaxy formation: stars that end up in a descendant galaxy were not necessarily formed in-situ. The most significant biases in stellar mass assembly histories are seen for lower redshift descendant galaxies at longer lookback times (see Figure \ref{fig:ratio_vs_lookbacktime}) and for galaxies hosted by the most massive halos (see Figure \ref{fig:separate_halomass}). The lookback time dependence is interesting in light of recent observational studies: \cite{Carnall2024} argued that, of their sample of four quiescent galaxies, the inferred stellar mass assembly history of the lowest redshift source (the same source studied by \citealt{Glazebrook2024}, at $z=3.19$) is most challenging to models.

\subsection{Implications for observational studies of massive quenched galaxies}
\begin{figure}
    \centering
    \includegraphics[width=1.0\linewidth]{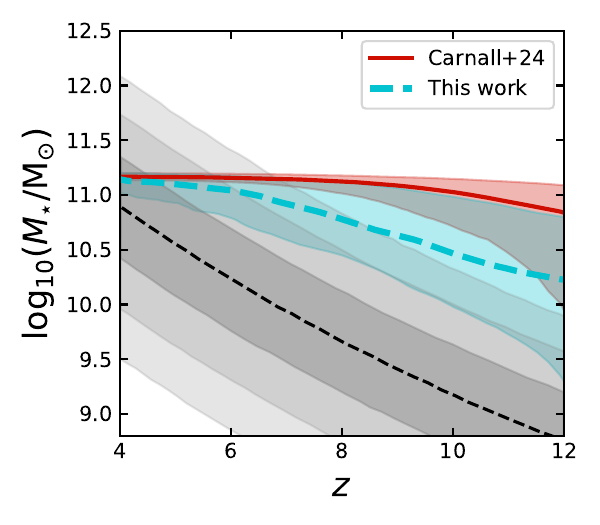}
    \caption{A demonstration of the application of our inferred bias to the stellar mass assembly history inferred for galaxy ZF-UDS-7329 by \protect\cite{Carnall2024}. The red line, inner and outer contours show median and  $1\sigma$ interval from their SED fit; this corresponds to $M_{\rm{archaeological}}$. I derive the equivalent blue line and interval by applying corrections derived for $z=3$ descendant galaxies in the $M_{\star}=10^{11}-10^{11.5}\,\rm{M_{\odot}}$ stellar mass range. The grey contours show expectations for stellar mass derived from Extreme Value Statistics (EVS), as calculated by \protect\cite{Carnall2024} using the fiducial star formation efficiency model outlined by \protect\cite{Lovell2023}. Tension with the EVS expectation is significantly alleviated after the likely contribution from ex-situ mass is accounted for.}\label{fig:Carnall_comparison}
\end{figure}

Our results have significant implications for the interpretation of spectroscopic observations of massive, high redshift galaxies. Fundamentally, past galaxy-galaxy mergers can significantly hamper the inference of the stellar mass assembly of an individual galaxy. This will also have an impact on the meaning of commonly-used quantities like formation redshift (i.e. the redshift at which a galaxy formed half its stellar mass) or quenching redshift/timescale \citep[see e.g.][]{Tacchella2022b}. The range of assembly histories, even for galaxies with similar stellar masses and redshifts, prevents very precise corrections for the effect of ex-situ mass assembly for individual sources. Nevertheless, here I study whether applying average corrections to the stellar assembly history of a particularly contentious source (ZF-UDS-7329; \citealt{Carnall2024,Glazebrook2024}) is sufficient to alleviate the claimed tension with $\Lambda\rm{CDM}$ or need for extreme star formation efficiencies at early times. In Figure \ref{fig:Carnall_comparison}, I reproduce the stellar assembly history of this source inferred from high-quality JWST NIRSpec data (red line, inner and outer contours show median, $1\sigma$ and $2\sigma$ intervals from their SED fit, respectively; see Figure 7 of \citealt{Carnall2024}). This contour was derived based on the stellar ages of all stars within the $z=3.19$ descendant, and can therefore be considered as $M_{\rm{archaeological}}$ versus $z$. I apply a correction based on that measured for simulated galaxies in the appropriate stellar mass range ($M_{\star}=10^{11}-10^{11.5}\,\rm{M_{\odot}}$) at $z=3$. In practice, I draw $1000$ samples from the posterior SFH of ZF-UDS-7329, as fitted by \cite{Carnall2024}. I multiply the SFH of each sample by the $M_{\rm{progenitor}}/M_{\rm{archaeological}}$ versus $z$ (i.e. similar to the inverse of the $z_{\rm{final}}=3$ panel of Figure \ref{fig:500_most_massive}), for every modelled galaxy in the chosen stellar mass range (about $60$ galaxies). This yields $1000\times60$ instances of the corrected SFH. I then characterise the distribution of $M_{\rm{progenitor}}$ at each redshift using these $60,000$ `corrected' samples. This yields an estimate of the stellar mass assembly history of the most massive progenitor (see blue line and contours). I overlay Extreme Value Statistics (EVS)-derived confidence intervals (grey contours), which were calculated by \cite{Carnall2024} using the tools provided by \cite{Lovell2023}, using their fiducial star formation efficiency model. It is clear that accounting for this bias has a significant impact on the tension between the inferred stellar mass history and the expected mass of the most massive galaxy within the survey volume. The revised median mass assembly history is now within $2-3\sigma$ expectations from the EVS PDF, out to $z\sim9$. This includes the $z=7-9$ range, where \cite{Carnall2024} noted a discrepancy between their model and EVS expectations. However, the EVS contours used in that work are averaged over cosmic environments and do not account for the large-scale fluctuations of the underlying dark matter density field, or for massive galaxies residing in the highest density regions. For these rare, massive sources found within relatively small fields, cosmic variance is significant \citep[see e.g.][]{Somerville2004,Moster2011}. As demonstrated by \cite{Jespersen2025}, conditioning the EVS contours based on massive galaxies being highly clustered and found in overdense environments raises expectations for stellar mass histories relative to the fiducial contours assumed by \cite{Carnall2024}, alleviating tension even further. At $z>9$, the inferred star formation history becomes very uncertain and, as noted by \cite{Carnall2024}, there is no tension with EVS expectations even before my correction. \\
\indent My results demonstrate that existing observations of massive quiescent galaxies neither challenge $\Lambda\rm{CDM}$ nor provide strong evidence for extremely high star formation efficiencies (i.e. near-100\% global conversion of baryons to stars, as suggested by \citealt{Carnall2024}) at early times. I note that there are many other uncertainties in SFH inference, in particular the degeneracies between formation time and metallicity discussed by \cite{DeGraaff2024} and the dependence on SFH prior demonstrated by \cite{Turner2025}, which prevent high-precision tests of cosmology. Nevertheless, the bias-corrected contours that I present in Figure \ref{fig:Carnall_comparison} still leave room for enhanced global star formation efficiency relative to the fiducial assumptions of \cite{Lovell2023}.\\
\indent My results are broadly in agreement with the conclusions drawn by \cite{Donnan2025}, who looked for evidence of high star formation efficiencies using a very different approach. \cite{Donnan2025} demonstrated that both the stellar mass function at $z=6-8$ (note that there, masses at, rather than prior to, the time of observation are inferred) and the UV luminosity function at $z=6-13$ can be fitted by a simple model with a non-evolving halo mass dependent star formation efficiency. They concluded that there is not yet strong evidence for increased star formation efficiency at early times (though note that this does require the decreasing typical age of the stellar population with redshift, as young as $\sim10\,\rm{Myr}$ at $z\sim13$).

\subsection{Challenges and avenues for future work}
As we saw from Figure \ref{fig:halo_hist}, individual galaxies show a range of merger histories. This makes it challenging to derive appropriate corrections for individual sources. Although applying an average correction to the inferred stellar mass assembly, as I have done in this paper, goes some way towards a more robust comparison with halo histories, individual sources may have particularly extreme merger growth histories. Larger samples of early quiescent galaxies like ZF-UDS-7329, for which stellar mass histories could be corrected on a population basis, might provide better tests for halo growth models.\\
\indent Interestingly, the archaeological stellar mass history inferred from the spectrum of ZF-UDS-7329 is very flat compared to the archaeological histories constructed from all progenitors of the massive $z=3$ galaxies in TNG-100 (see Figure \ref{fig:halo_hist}). Although there are large uncertainties on the derived star formation history of ZF-UDS-7329, as well as significant degeneracies with metallicity, this hints that the TNG model may not form massive galaxies as early as required by observations. This likely reflects limitations of the subgrid models employed to model star formation and feedback, and the simplistic treatment of black hole feedback, alongside tuning to match $z\sim0$ data. This can be addressed in comparisons between measured and predicted stellar mass functions (ideally, separated into star-forming and quiescent sources, e.g. \citealt{McLeod2021,Weaver2023}). Recent comparisons between modelled and measured stellar mass functions do not find significant tension \citep[see][]{Weibel2024a}, but there is considerable variation between different models, and only upper limits can be placed on stellar mass functions for $M_{\star}\gtrsim10^{10}\,\rm{M_{\odot}}$ at $z\gtrsim6$. In the case of ZF-UDS-7329, mergers between already-quenched galaxies may be required to explain the flat star formation history; this may be hinted at by the larger inferred stellar size of ZF-UDS-7329 compared to other high-redshift quiescent galaxies (see \citealt{Kawinwanichakij2025}, for a compilation). \cite{Turner2025} investigated colour gradients within ZF-UDS-7329, probing the rest-frame $U$, $V$ and $J$ bands with NIRCam imaging in the F150W, F200W, and F444W bands, respectively. They found a younger, bluer disk-like component with inferred stellar age $<0.5\,\rm{Gyr}$, alongside a significantly older, redder, bulge-like component that was probed by NIRSpec. They argued that the combination of a relatively large size ($>1\,\rm{kpc}$) and a colour gradient was consistent with the galaxy having undergone previous merger(s). Future work could also explore the formation histories of massive early-quenching galaxies within the semi-analytic model calibration framework presented by \cite{Araya-Araya2025}, who calibrated the \texttt{L-Galaxies} semi-analytic model using number densities of quenched galaxies as well as sub-millimetre number counts. \\
\indent \cite{Baker2025a} study the star formation histories of massive quiescent galaxies within the FLAMINGO simulation, motivated by its large size $(1\,\rm{cGpc})^{3}$. They find similar number densities of massive quiescent galaxies to TNG300 at $z\lesssim4$. \cite{Baker2025a} study the in-situ versus ex-situ mass fractions as a function of time, finding that the fraction of stellar mass formed ex-situ is generally $\lesssim40\%$. Although this is a slightly different calculation to the one performed in this paper, my convergence tests may have implications for this result. In Figures \ref{fig:convergence_test} and \ref{fig:convergence_test_tng300}, I compare $M_{\star,\,\rm{archaeological}}/M_{\star,\,\rm{progenitor}}$ versus stellar mass at different lookback times, for TNG50, TNG100 and TNG300. While TNG100 shows good agreement with TNG50, TNG300 shows significant disagreement with TNG100, which is worse at the lowest stellar masses. This indicates that the lower mass resolution of TNG300 is insufficient to quantify $M_{\star,\,\rm{archaeological}}/M_{\star,\,\rm{progenitor}}$ at early times. For the FLAMINGO box used in the work of \cite{Baker2025a}, the baryonic particle mass is $1.34\times10^{8}\,\rm{M_{\odot}}$, i.e. over $10$ times lower resolution than TNG300. Hence, mergers of low-mass galaxies will not be resolved, and the ex-situ mass assembly is likely underestimated.

\section{Conclusions}\label{sec:conclusions}
In this work, I use the IllustrisTNG100 simulation to study the assembly histories of massive galaxies at different cosmic epochs. For each massive galaxy, I calculate the `archaeological' stellar mass assembly using the ages of all stars present at the time of observation. This corresponds to the ideal-case stellar mass history that can be constructed from spectroscopic observations. I compare this archaeological mass history to the stellar mass contained in the single most massive progenitor, as a function of lookback time. I find that these two stellar mass tracks can be highly discrepant, as a result of mass being formed ex-situ at early times. For example, for a massive ($M_{\star}\gtrsim 2\times10^{11}\,\rm{M_{\odot}}$) galaxy observed at $z=1$ ($z=2$), the stellar mass of an individual progenitor is, on average, overestimated by a factor of $\gtrsim3$ ($\gtrsim2$) above $z=6$. The discrepancy between archaeological and most-massive-progenitor mass increases for sources observed at lower redshift, with lookback time, and with halo mass. However, there are large galaxy-to-galaxy variations in merger histories. This prevents detailed comparison of past stellar mass with halo mass models. For a $M_{\star}\sim10^{11}\,\rm{M_{\odot}}$ galaxy at $z\sim3$, like the massive, quenched observed source ZF-UDS-7329, the typical bias is a factor of $\sim3$ at $z=8$. Correcting the inferred stellar mass history for this bias reduces tensions with halo growth models that have been claimed in recent work. I conclude that there is no significant evidence for a challenge to $\Lambda\rm{CDM}$, nor do the data demand extreme star formation efficiencies at early times. \\
\indent Although this work was motivated by recent studies of high-redshift quiescent galaxies, the implications of my findings are not limited to quiescent galaxies. In general, the hierarchical nature of galaxy assembly leads to significant uncertainties in the inference of stellar mass histories that have not previously been accounted for. I provide TNG100-based typical corrections that may be useful for incorporating this effect into future work. 

\section*{Acknowledgements}
RKC thanks the reviewer, Sandro Tacchella, for a thorough reading of the paper and helpful suggestions. RKC is grateful for support from the Leverhulme Trust via the Leverhulme Early Career Fellowship. RKC thanks Jim Dunlop and Ross McLure for helpful comments on the manuscript, as well as Daniel Angl\'{e}s-Alc\'{a}zar, Callum Donnan, and Sandro Tacchella for productive conversations. RKC thanks Adam Carnall for sharing the SED fit of ZF-UDS-7329.
\section*{Data Availability}
TNG data and accompanying script examples are available here: \url{https://www.tng-project.org/data/}. In this paper, I use the TNG100 SubLink merger trees, available here: \url{https://www.tng-project.org/data/downloads/TNG100-1/}.
\bibliographystyle{mnras}
\bibliography{Edinburgh} 

\begin{thebibliography}{}
\makeatletter
\relax
\def\mn@urlcharsother{\let\do\@makeother \do\$\do\&\do\#\do\^\do\_\do\%\do\~}
\def\mn@doi{\begingroup\mn@urlcharsother \@ifnextchar [ {\mn@doi@} {\mn@doi@[]}}
\def\mn@doi@[#1]#2{\def\@tempa{#1}\ifx\@tempa\@empty \href {http://dx.doi.org/#2} {doi:#2}\else \href {http://dx.doi.org/#2} {#1}\fi \endgroup}
\def\mn@eprint#1#2{\mn@eprint@#1:#2::\@nil}
\def\mn@eprint@arXiv#1{\href {http://arxiv.org/abs/#1} {{\tt arXiv:#1}}}
\def\mn@eprint@dblp#1{\href {http://dblp.uni-trier.de/rec/bibtex/#1.xml} {dblp:#1}}
\def\mn@eprint@#1:#2:#3:#4\@nil{\def\@tempa {#1}\def\@tempb {#2}\def\@tempc {#3}\ifx \@tempc \@empty \let \@tempc \@tempb \let \@tempb \@tempa \fi \ifx \@tempb \@empty \def\@tempb {arXiv}\fi \@ifundefined {mn@eprint@\@tempb}{\@tempb:\@tempc}{\expandafter \expandafter \csname mn@eprint@\@tempb\endcsname \expandafter{\@tempc}}}

\bibitem[\protect\citeauthoryear{Antwi-Danso et~al.,}{Antwi-Danso et~al.}{2025}]{Antwi-Danso2025}
Antwi-Danso J.,  et~al., 2025, \mn@doi [ApJ] {10.3847/1538-4357/ad8b30}, 978, 90

\bibitem[\protect\citeauthoryear{Araya-Araya, Cochrane, Hayward, Sodré, Yates, Daalen  \& Vicentin}{Araya-Araya et~al.}{2025}]{Araya-Araya2025}
Araya-Araya P.,  Cochrane R.,  Hayward C.~C.,  Sodré L.,  Yates R.~M.,  Daalen M.~V.,   Vicentin M.,  2025, arXiv:2504.15283

\bibitem[\protect\citeauthoryear{Avila et~al.,}{Avila et~al.}{2014}]{Avila2014}
Avila S.,  et~al., 2014, \mn@doi [MNRAS] {10.1093/mnras/stu799}, 441, 3488

\bibitem[\protect\citeauthoryear{Baker et~al.,}{Baker et~al.}{2025}]{Baker2025a}
Baker W.~M.,  et~al., 2025, \mn@doi [MNRAS] {10.1093/mnras/staf475}, 539, 557

\bibitem[\protect\citeauthoryear{Baugh, Cole, Frenk  \& Lacey}{Baugh et~al.}{1998}]{Baugh1998}
Baugh C.~M.,  Cole S.,  Frenk C.~S.,   Lacey C.~G.,  1998, \mn@doi [ApJ] {10.1007/BF02709321}, 498, 504

\bibitem[\protect\citeauthoryear{Baugh, Lacey, Frenk, Granato, Silva, Bressan, Benson  \& Cole}{Baugh et~al.}{2005}]{Baugh2005}
Baugh C.~M.,  Lacey C.~G.,  Frenk C.~S.,  Granato G.~L.,  Silva L.,  Bressan A.,  Benson A.~J.,   Cole S.,  2005, \mn@doi [MNRAS] {10.1111/j.1365-2966.2004.08553.x}, 356, 1191

\bibitem[\protect\citeauthoryear{Blumenthal, Faber, Rees  \& Primack}{Blumenthal et~al.}{1984}]{Blumenthal1984}
Blumenthal G.~R.,  Faber S.~M.,  Rees M.~J.,   Primack J.~R.,  1984, Nature, 311, 517

\bibitem[\protect\citeauthoryear{Bond, Cole, Efstathiou  \& Kaiser}{Bond et~al.}{1991}]{Bond1991}
Bond J.~R.,  Cole S.,  Efstathiou G.,   Kaiser N.,  1991, \mn@doi [ApJ] {10.1086/170520}, 379, 440

\bibitem[\protect\citeauthoryear{Bower}{Bower}{1991}]{Bower1991}
Bower R.~G.,  1991, MNRAS, 248, 332

\bibitem[\protect\citeauthoryear{Bower, a. J~Benson, Malbon, Helly, Frenk, Baugh, Cole  \& Lacey}{Bower et~al.}{2006}]{Bower2006}
Bower R.~G.,  a. J~Benson Malbon R.,  Helly J.~C.,  Frenk C.~S.,  Baugh C.~M.,  Cole S.,   Lacey C.~G.,  2006, \mn@doi [MNRAS] {10.1111/j.1365-2966.2006.10519.x}, 370, 645

\bibitem[\protect\citeauthoryear{Brinchmann \& Ellis}{Brinchmann \& Ellis}{2000}]{Brinchmann2000}
Brinchmann J.,  Ellis R.~S.,  2000, ApJ, 536, 77

\bibitem[\protect\citeauthoryear{Bundy et~al.,}{Bundy et~al.}{2006}]{Bundy2006}
Bundy K.,  et~al., 2006, ApJ, 651, 120

\bibitem[\protect\citeauthoryear{Carnall et~al.,}{Carnall et~al.}{2023a}]{Carnall2023}
Carnall A.~C.,  et~al., 2023a, \mn@doi [MNRAS] {10.1093/mnras/stad369}, 520, 3974

\bibitem[\protect\citeauthoryear{Carnall et~al.,}{Carnall et~al.}{2023b}]{Carnall2023a}
Carnall A.~C.,  et~al., 2023b, \mn@doi [Nature] {10.1038/s41586-023-06158-6}, 619, 716

\bibitem[\protect\citeauthoryear{Carnall et~al.,}{Carnall et~al.}{2024}]{Carnall2024}
Carnall A.~C.,  et~al., 2024, \mn@doi [MNRAS] {10.1093/mnras/stae2092}, 534, 325

\bibitem[\protect\citeauthoryear{Cowie, Songaila, Hu  \& Cohen}{Cowie et~al.}{1996}]{Cowie1996}
Cowie L.,  Songaila A.,  Hu E.,   Cohen J.,  1996, ApJ, 112, 839

\bibitem[\protect\citeauthoryear{Croton et~al.,}{Croton et~al.}{2005}]{Croton2005}
Croton D.~J.,  et~al., 2005, MNRAS, 356, 1155

\bibitem[\protect\citeauthoryear{Davis, Efstathiou, Frenk  \& White}{Davis et~al.}{1985}]{Davis1985}
Davis M.,  Efstathiou G.,  Frenk C.~S.,   White S. D.~M.,  1985, ApJ, 292, 371

\bibitem[\protect\citeauthoryear{Dolag, Borgani, Murante  \& Springel}{Dolag et~al.}{2009}]{Dolag2009}
Dolag K.,  Borgani S.,  Murante G.,   Springel V.,  2009, \mn@doi [MNRAS] {10.1111/j.1365-2966.2009.15034.x}, 399, 497

\bibitem[\protect\citeauthoryear{Donnan, Dunlop, McLure, McLeod  \& Cullen}{Donnan et~al.}{2025}]{Donnan2025}
Donnan C.~T.,  Dunlop J.~S.,  McLure R.~J.,  McLeod D.~J.,   Cullen F.,  2025, MNRAS, 539, 2409

\bibitem[\protect\citeauthoryear{Duan et~al.,}{Duan et~al.}{2025}]{Duan2025}
Duan Q.,  et~al., 2025, \mn@doi [MNRAS] {10.1093/mnras/staf638}, 540, 774

\bibitem[\protect\citeauthoryear{Genel et~al.,}{Genel et~al.}{2014}]{Genel2014}
Genel S.,  et~al., 2014, \mn@doi [MNRAS] {10.1093/mnras/stu1654}, 445, 175

\bibitem[\protect\citeauthoryear{Glazebrook et~al.,}{Glazebrook et~al.}{2004}]{Glazebrook2004}
Glazebrook K.,  et~al., 2004, \mn@doi [Nature] {10.1038/nature02667}, 430, 181

\bibitem[\protect\citeauthoryear{Glazebrook et~al.,}{Glazebrook et~al.}{2024}]{Glazebrook2024}
Glazebrook K.,  et~al., 2024, \mn@doi [Nature] {10.1038/s41586-024-07191-9}, 628, 277

\bibitem[\protect\citeauthoryear{Jespersen, Carnall  \& Lovell}{Jespersen et~al.}{2025}]{Jespersen2025}
Jespersen C.~K.,  Carnall A.~C.,   Lovell C.~C.,  2025, \mn@doi [ApJL] {10.3847/2041-8213/adeb7c}, 988, L19

\bibitem[\protect\citeauthoryear{Kauffmann, Colberg, Diaferio  \& White}{Kauffmann et~al.}{1999a}]{Kauffmann1999}
Kauffmann G.,  Colberg J.~M.,  Diaferio A.,   White S.~D.,  1999a, \mn@doi [MNRAS] {10.1046/j.1365-8711.1999.02202.x}, 303, 188

\bibitem[\protect\citeauthoryear{Kauffmann, Colberg, Diaferio  \& White}{Kauffmann et~al.}{1999b}]{Kauffmann1999a}
Kauffmann G.,  Colberg J.~M.,  Diaferio A.,   White S.~D.,  1999b, \mn@doi [MNRAS] {10.1046/j.1365-8711.1999.02202.x}, 307, 529

\bibitem[\protect\citeauthoryear{Kawinwanichakij et~al.,}{Kawinwanichakij et~al.}{2025}]{Kawinwanichakij2025}
Kawinwanichakij L.,  et~al., 2025, \mn@doi [arXiv:2505.03089] {10.48550/arXiv.2505.03089}

\bibitem[\protect\citeauthoryear{Kriek et~al.,}{Kriek et~al.}{2018}]{2018ascl.soft03008K}
Kriek M.,  et~al., 2018, FAST: Fitting and Assessment of Synthetic Templates

\bibitem[\protect\citeauthoryear{Lacey \& Cole}{Lacey \& Cole}{1993}]{Lacey1993}
Lacey C.,  Cole S.,  1993, \mn@doi [MNRAS] {10.1093/mnras/262.3.627}, 262, 627

\bibitem[\protect\citeauthoryear{Lacey \& Cole}{Lacey \& Cole}{1994}]{Lacey1994}
Lacey C.,  Cole S.,  1994, \mn@doi [MNRAS] {10.1093/mnras/271.3.676}, 271, 676

\bibitem[\protect\citeauthoryear{Lee et~al.,}{Lee et~al.}{2014}]{Lee2014}
Lee J.,  et~al., 2014, \mn@doi [MNRAS] {10.1093/mnras/stu2039}, 445, 4197

\bibitem[\protect\citeauthoryear{Lovell, Harrison, Harikane, Tacchella  \& Wilkins}{Lovell et~al.}{2023}]{Lovell2023}
Lovell C.~C.,  Harrison I.,  Harikane Y.,  Tacchella S.,   Wilkins S.~M.,  2023, \mn@doi [MNRAS] {10.1093/mnras/stac3224}, 518, 2511

\bibitem[\protect\citeauthoryear{Lucia \& Blaizot}{Lucia \& Blaizot}{2007}]{DeLucia2007}
Lucia G.~D.,  Blaizot J.,  2007, \mn@doi [MNRAS] {10.1111/j.1365-2966.2006.11287.x}, 375, 2

\bibitem[\protect\citeauthoryear{Lucia, Springel, White, Croton  \& Kauffmann}{Lucia et~al.}{2006}]{DeLucia2006}
Lucia G.~D.,  Springel V.,  White S.~D.,  Croton D.,   Kauffmann G.,  2006, \mn@doi [MNRAS] {10.1111/j.1365-2966.2005.09879.x}, 366, 499

\bibitem[\protect\citeauthoryear{Marinacci et~al.,}{Marinacci et~al.}{2018}]{Marinacci2018}
Marinacci F.,  et~al., 2018, \mn@doi [MNRAS] {10.1093/mnras/sty2206}, 480, 5113

\bibitem[\protect\citeauthoryear{McLeod, McLure, Dunlop, Cullen, Carnall  \& Duncan}{McLeod et~al.}{2021}]{McLeod2021}
McLeod D.~J.,  McLure R.~J.,  Dunlop J.~S.,  Cullen F.,  Carnall A.~C.,   Duncan K.,  2021, \mn@doi [MNRAS] {10.1093/mnras/stab731}, 503, 4413

\bibitem[\protect\citeauthoryear{Moster, Somerville, Newman  \& Rix}{Moster et~al.}{2011}]{Moster2011}
Moster B.~P.,  Somerville R.~S.,  Newman J.~A.,   Rix H.~W.,  2011, \mn@doi [ApJ] {10.1088/0004-637X/731/2/113}, 731

\bibitem[\protect\citeauthoryear{Naiman et~al.,}{Naiman et~al.}{2018}]{Naiman2018}
Naiman J.~P.,  et~al., 2018, \mn@doi [MNRAS] {10.1093/mnras/sty618}, 477, 1206

\bibitem[\protect\citeauthoryear{Nanayakkara et~al.,}{Nanayakkara et~al.}{2024}]{Nanayakkara2024}
Nanayakkara T.,  et~al., 2024, \mn@doi [Sci Rep] {10.1038/s41598-024-52585-4}, 14, 3724

\bibitem[\protect\citeauthoryear{Nanayakkara et~al.,}{Nanayakkara et~al.}{2025}]{Nanayakkara2024a}
Nanayakkara T.,  et~al., 2025, ApJ, 981, 78

\bibitem[\protect\citeauthoryear{Nelson et~al.,}{Nelson et~al.}{2018}]{Nelson2018a}
Nelson D.,  et~al., 2018, \mn@doi [MNRAS] {10.1093/mnras/stx3040}, 475, 624

\bibitem[\protect\citeauthoryear{Peacock \& Heavens}{Peacock \& Heavens}{1990}]{Peacock1990}
Peacock J.~A.,  Heavens A.~F.,  1990, MNRAS, 243, 133

\bibitem[\protect\citeauthoryear{Peebles}{Peebles}{1982}]{Peebles1982}
Peebles P.,  1982, \mn@doi [ApJ] {10.1017/CBO9781107415324.004}, 263, L1

\bibitem[\protect\citeauthoryear{Peebles \& Yu}{Peebles \& Yu}{1970}]{Peebles1970}
Peebles P.,  Yu J.,  1970, ApJ, 162, 815

\bibitem[\protect\citeauthoryear{Pillepich et~al.,}{Pillepich et~al.}{2018}]{Pillepich2018}
Pillepich A.,  et~al., 2018, \mn@doi [MNRAS] {10.1093/mnras/stx3112}, 475, 648

\bibitem[\protect\citeauthoryear{{Planck Collaboration}}{{Planck Collaboration}}{2016}]{PlanckCollaboration2015}
{Planck Collaboration} 2016, A\&A, 594, A13

\bibitem[\protect\citeauthoryear{Press \& Schechter}{Press \& Schechter}{1974}]{Press1974}
Press W.~H.,  Schechter P.,  1974, \mn@doi [ApJ] {10.1086/152650}, 187, 425

\bibitem[\protect\citeauthoryear{Puskás et~al.,}{Puskás et~al.}{2025}]{Puskas2025}
Puskás D.,  et~al., 2025, \mn@doi [MNRAS] {10.1093/mnras/staf813}, 540, 2146

\bibitem[\protect\citeauthoryear{Renzini}{Renzini}{2006}]{Renzini2006}
Renzini A.,  2006, \mn@doi [ARAA] {10.1146/annurev.astro.44.051905.092450}, 44, 141

\bibitem[\protect\citeauthoryear{Rodriguez-Gomez et~al.,}{Rodriguez-Gomez et~al.}{2015}]{Rodriguez-Gomez2015}
Rodriguez-Gomez V.,  et~al., 2015, \mn@doi [MNRAS] {10.1093/mnras/stv264}, 449, 49

\bibitem[\protect\citeauthoryear{Schreiber, Elbaz, Pannella, Ciesla, Wang  \& Franco}{Schreiber et~al.}{2018}]{Schreiber2018}
Schreiber C.,  Elbaz D.,  Pannella M.,  Ciesla L.,  Wang T.,   Franco M.,  2018, \mn@doi [A\\&A] {10.1051/0004-6361/201731506}, 609, 1

\bibitem[\protect\citeauthoryear{Setton et~al.,}{Setton et~al.}{2024}]{Setton2024a}
Setton D.~J.,  et~al., 2024, \mn@doi [ApJ] {10.3847/1538-4357/ad6a18}, 974, 145

\bibitem[\protect\citeauthoryear{Sijacki, Vogelsberger, Genel, Springel, Torrey, Snyder, Nelson  \& Hernquist}{Sijacki et~al.}{2015}]{Sijacki2015}
Sijacki D.,  Vogelsberger M.,  Genel S.,  Springel V.,  Torrey P.,  Snyder G.~F.,  Nelson D.,   Hernquist L.,  2015, \mn@doi [MNRAS] {10.1093/mnras/stv1340}, 452, 575

\bibitem[\protect\citeauthoryear{Somerville, Lee, Ferguson, Gardner, Moustakas  \& Giavalisco}{Somerville et~al.}{2004}]{Somerville2004}
Somerville R.~S.,  Lee K.,  Ferguson H.~C.,  Gardner J.~P.,  Moustakas L.~A.,   Giavalisco M.,  2004, ApJ, 600, L171

\bibitem[\protect\citeauthoryear{Springel}{Springel}{2010}]{Springel2010}
Springel V.,  2010, \mn@doi [MNRAS] {10.1111/j.1365-2966.2009.15715.x}, 401, 791

\bibitem[\protect\citeauthoryear{Springel, White, Tormen  \& Kauffmann}{Springel et~al.}{2001}]{Springel2001}
Springel V.,  White S.~D.,  Tormen G.,   Kauffmann G.,  2001, \mn@doi [MNRAS] {10.1086/318381}, 328, 726

\bibitem[\protect\citeauthoryear{Springel et~al.,}{Springel et~al.}{2018}]{Springel2018}
Springel V.,  et~al., 2018, \mn@doi [MNRAS] {10.1093/mnras/stx3304}, 475, 676

\bibitem[\protect\citeauthoryear{Srisawat et~al.,}{Srisawat et~al.}{2013}]{Srisawat2013}
Srisawat C.,  et~al., 2013, \mn@doi [MNRAS] {10.1093/mnras/stt1545}, 436, 150

\bibitem[\protect\citeauthoryear{Tacchella et~al.,}{Tacchella et~al.}{2022}]{Tacchella2022b}
Tacchella S.,  et~al., 2022, \mn@doi [ApJ] {10.3847/1538-4357/ac449b}, 926, 134

\bibitem[\protect\citeauthoryear{Turner et~al.,}{Turner et~al.}{2025}]{Turner2025}
Turner C.,  et~al., 2025, \mn@doi [MNRAS] {10.1093/mnras/staf475}, pp 1826--1848

\bibitem[\protect\citeauthoryear{Vogelsberger et~al.,}{Vogelsberger et~al.}{2014a}]{Vogelsberger2014a}
Vogelsberger M.,  et~al., 2014a, \mn@doi [MNRAS] {10.1093/mnras/stu1536}, 444, 1518

\bibitem[\protect\citeauthoryear{Vogelsberger et~al.,}{Vogelsberger et~al.}{2014b}]{Vogelsberger2014}
Vogelsberger M.,  et~al., 2014b, \mn@doi [Nature] {10.1038/nature13316}, 509, 177

\bibitem[\protect\citeauthoryear{Weaver et~al.,}{Weaver et~al.}{2023}]{Weaver2023}
Weaver J.~R.,  et~al., 2023, \mn@doi [A\\&A] {10.1051/0004-6361/202245581}, 677, A184

\bibitem[\protect\citeauthoryear{Weibel et~al.,}{Weibel et~al.}{2024}]{Weibel2024a}
Weibel A.,  et~al., 2024, \mn@doi [MNRAS] {10.1093/mnras/stae1891}, 533, 1808

\bibitem[\protect\citeauthoryear{Weinberger, Springel  \& Pakmor}{Weinberger et~al.}{2020}]{Weinberger2020}
Weinberger R.,  Springel V.,   Pakmor R.,  2020, \mn@doi [The Astrophysical Journal Supplement Series] {10.3847/1538-4365/ab908c}, 248, 32

\bibitem[\protect\citeauthoryear{White \& Rees}{White \& Rees}{1978}]{White1978}
White S. D.~M.,  Rees M.~J.,  1978, MNRAS, 183, 341

\bibitem[\protect\citeauthoryear{de Graaff et~al.,}{de~Graaff et~al.}{2024}]{DeGraaff2024}
de Graaff A.,  et~al., 2024, \mn@doi [Nature Astronomy] {10.1038/s41550-024-02424-3}

\bibitem[\protect\citeauthoryear{van~der Wel, Franx, van Dokkum, Rix, Illingworth  \& Rosati}{van~der Wel et~al.}{2005}]{vanderWel2005}
van~der Wel A.,  Franx M.,  van Dokkum P.~G.,  Rix H.,  Illingworth G.~D.,   Rosati P.,  2005, \mn@doi [ApJ] {10.1086/430464}, 631, 145

\makeatother
\end{thebibliography}


\appendix
\section{Resolution tests}\label{sec:appendix_res_tests}
\begin{figure}
    \centering
    \includegraphics[width=1.0\linewidth]{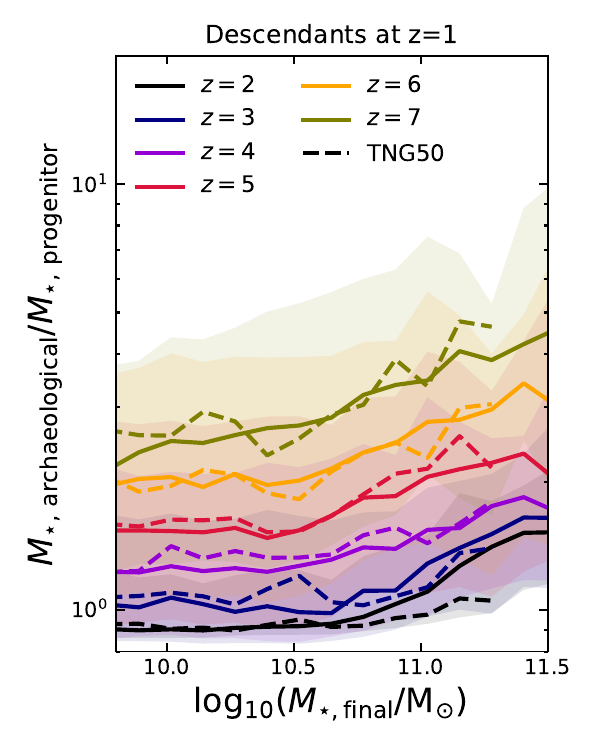}
    \caption{An expanded version of the right-hand panel of Figure \ref{fig:mass_dependence} (descendants at $z=1$), with additional dashed lines showing equivalent relation derived using TNG50. At all lookback times, the median relation derived using TNG50 falls within the $1\sigma$ contours derived from TNG100. This indicates that my results are insensitive to the specific TNG model used, and that TNG100 has sufficient resolution for these tests. Note that there is increased noise in the TNG50 relations due to having fewer galaxies in the smaller box.}
    \label{fig:convergence_test}
\end{figure}

\begin{figure}
    \centering
    \includegraphics[width=1.0\linewidth]{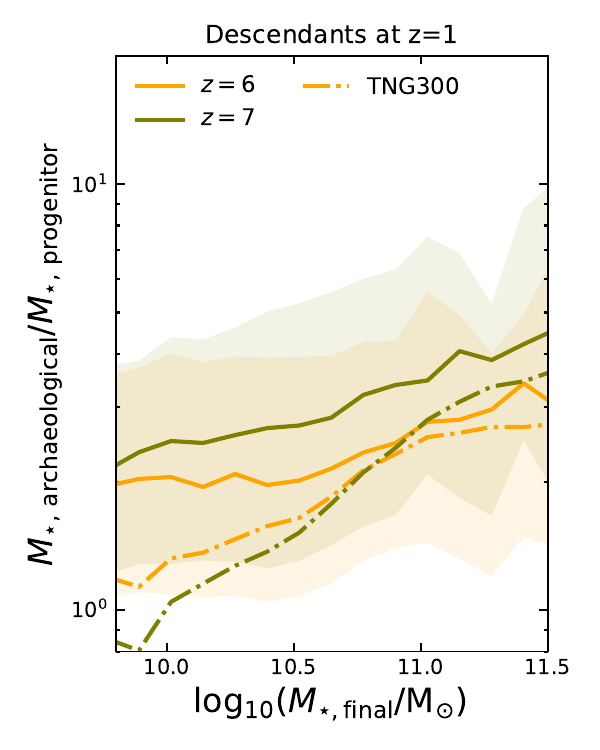}
    \caption{As Figure \ref{fig:convergence_test}, but comparing results from the fiducial model (TNG100, solid lines) to the lower resolution TNG300 (dashed lines). For clarity, I show only two lookback times, $z=6$ and $z=7$. There is clearly a lack of convergence in this relation for TNG300, which worsens for lower stellar mass galaxies. This highlights the importance of particle resolution for studies of in-situ versus ex-situ mass assembly.}
    \label{fig:convergence_test_tng300}
\end{figure}

I confirm that TNG100 has sufficient resolution for this study by repeating the analyses at $z=1$ using the higher resolution TNG50 simulation. In the TNG50 box, dark matter particles have mass $4.5\times10^{5}\,\rm{M_{\odot}}$ and baryons have mass $8.5\times10^{4}\,\rm{M_{\odot}}$. These masses are over $10$ times lower than the equivalent masses in TNG100. I summarise my results in Figure \ref{fig:convergence_test}. At fixed stellar mass, for all lookback times (at $z=2-8$), the median $M_{\star,\,\rm{archaeological}}/M_{\star,\,\rm{progenitor}}$ calculated with TNG50 is within the $1\sigma$ contours calculated with TNG100. This indicates that my results are not being affected by the relatively low resolution of TNG100.\\
\indent I also explore the impact of a much lower resolution simulation on the results. In Figure \ref{fig:convergence_test_tng300}, I compare the median $M_{\star,\,\rm{archaeological}}/M_{\star,\,\rm{progenitor}}$ predicted by TNG100-1 and TN300-1 at $z=6$ and $z=7$. The predictions of TNG300 are significantly lower, with an offset that increases towards low stellar mass. This reflects the lower resolution of TNG300 (baryonic mass resolution $1.1\times10^{7}\,\rm{M_{\odot}}$) compared to TNG100 (baryonic mass resolution $1.4\times10^{6}\,\rm{M_{\odot}}$) and highlights the importance of convergence testing to ensure that physical conclusions are not driven by resolution effects.

\bsp	
\label{lastpage}
\end{document}